\documentclass[final,twocolumn]{elsarticle}

\usepackage{amssymb}

\begin{document}

\begin{frontmatter}



\title{``Pudding Mold''-type Band as an Origin of Large Thermopower 
in $\tau$-type Organic Conductors}


\author[label1]{Hirohito Aizawa}
\author{Kazuhiko Kuroki\corref{cor1}\fnref{label3}}
\ead{kuroki@vivace.e-one.uec.ac.jp}
\cortext[cor1]{Corresponding author.}
\author[label2]{Harukazu Yoshino}
\author[label2]{Keizo Murata}

\address[label1]{ Department of Applied Physics and Chemistry, 
 The University of Electro-Communications, Chofu, Tokyo 182-8585, Japan}
\address[label2]{ 
Graduate School of Science, Osaka City University, Osaka 558-8585, Japan}

\begin{abstract}

We study the origin of the large thermopower in quasi-two-dimensional 
a $\tau$-type organic conductor, 
$\tau$-(EDO-\textit{S,S}-DMEDT-TTF)$_{2}$(AuBr$_{2}$)$_{1+y}$ 
($y \le 0.875$), 
from the view point of a ``pudding mold''-type band structure. 
We calculate the electronic band structure using 
an \textit{ab initio} band calculation package, 
and obtain a tight binding model fit to 
the \textit{ab initio} band structure. 
Using the model and the Boltzmann's equation approach, 
we calculate the temperature dependence of the Seebeck coefficient. 
We conclude that the peculiar band structure is the 
origin of the large Seebeck coefficient and the appearance of the 
maximum value at a certain temperature.

\end{abstract}


\end{frontmatter}



\section{Introduction} 
\label{Introduction}

Recently, large thermopower has been 
experimentally observed in quasi-two-dimensional organic conductors 
having $\tau$-type alignment of the donor molecules, 
e. g. $\tau$-(EDO-\textit{S,S}-DMEDT-TTF)$_{2}$(AuBr$_{2}$)$_{1+y}$ 
($y \le 0.875$)\cite{Yoshino-Papavassiliou-JTAC}. 
The absolute value of the Seebeck coefficient takes its maximum 
of about $\left| S \right| \simeq 150$ $\mu$V/K 
around the temperature of $T \simeq 100$ K to $T \simeq 150$ K. 

A possible link between the large thermopower and 
the electronic structure is of special interest. 
In fact, a calculated band structure for 
$\tau$-(EDO-\textit{S,S}-DMEDT-TTF)$_{2}$(AuBr$_{2}$)$_{1+y}$ 
using the extended H$\ddot{\rm u}$ckel method 
\cite{Papavassiliou-Lagouvardos-MCLC-285-83} 
has shown presence of a star-shaped Fermi surface and a peculiar 
band structure with flat portions near the $\Gamma$ point. 
Another feature of this material 
is the variation of the  band-filling corresponding to the 
amount of anions. The anion content of $y$ corresponds to the 
electron band filling of $n=1.5-y/2$.

In the present study, 
we perform an \textit{ab initio} band calculation 
for $\tau$-(EDO-\textit{S,S}-DMEDT-TTF)$_{2}$AuBr$_{2}$ using 
the WIEN2K package,\cite{WIEN2K} 
and make a tight binding model fit to the obtained \textit{ab initio} 
band structure. 
We calculate the Seebeck coefficient using the Boltzmann's equation 
approach and show that 
the peculiar band structure is the origin of 
the large Seebeck coefficient.

\section{Band structure} 
\label{Band structure} 

We have performed calculations using 
all-electron full potential linearized augmented plane-wave (LAPW) 
+ local orbitals (lo) method to solve the Kohn-Sham equations 
using density functional theory (DFT) 
within the framework of WIEN2K \cite{WIEN2K}.
This implements the DFT with different possible approximation 
for the exchange correlation potentials. 
The exchange correlation potential is calculated using 
the generalized gradient approximation (GGA). 
The single-particle wave functions in the interstitial region are 
expanded by plane waves with a cutoff of 
$R_{\rm MT} K_{\rm max}=3.0$, 
where $R_{\rm MT}$ denotes the smallest muffin tin radius 
and $K_{\rm max}$ is the maximum value of the $K$ vector 
in the plane wave expansion. 
In $\tau$-(EDO-\textit{S,S}-DMEDT-TTF)$_{2}$AuBr$_{2}$, 
the muffin-tin radii are assumed to 
be 2.38, 2.11, 1.61, 1.27, 1.18 and 0.64 atomic units (au) 
for  Au,   Br,    S,    O,    C and    H, respectively. 
For the value of $K_{\rm max}$ and the plane wave cutoff energy, 
$K_{\rm max}=4.7$ and 
the plane wave cutoff energy is 298.8 eV 
for $\tau$-(EDO-\textit{S,S}-DMEDT-TTF)$_{2}$AuBr$_{2}$. 
The wave functions in the muffin tin spheres are expanded up 
to $l_{\rm max} = 10$, 
while the charge density are Fourier expanded up to 
$G_{\rm max}=20$. 
Calculations are performed by using 512 $k$-points 
in the irreducible Brillouin zone. 

\begin{figure}[h]
\begin{center}
  \includegraphics[width=7.0cm]{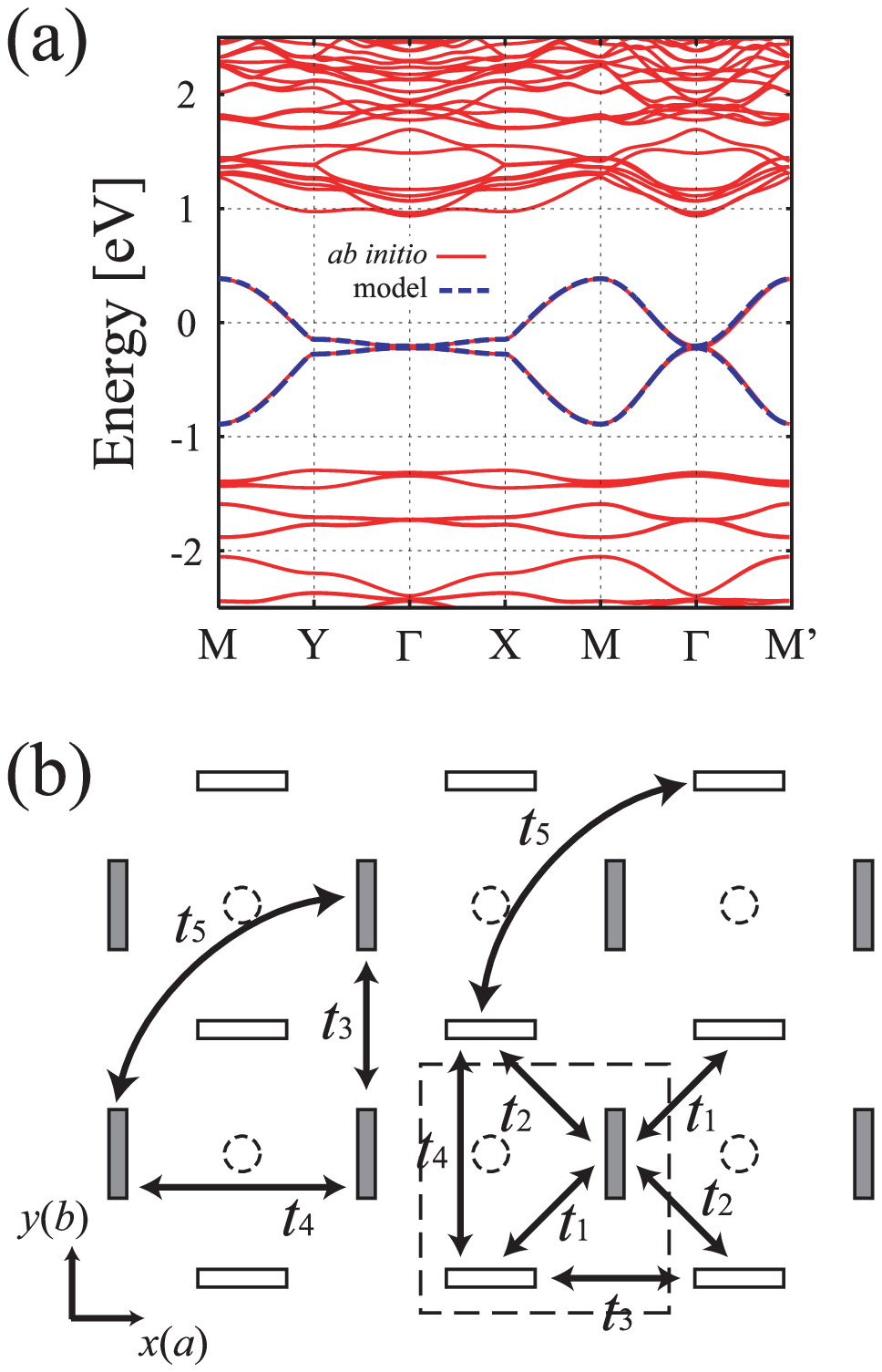}
\end{center}
{Fig.1
  (a) The band structure for the \textit{ab initio} result 
  (red solid lines) and the tight binding fit (blue dashed curves). 
  (b) The tight binding model in this study. 
  \label{fig1}}
 \end{figure}

The calculated band structure is shown in Figure \ref{fig1} (a). 
We have made a tight binding model fit to the \textit{ab initio} band 
as shown in Figure \ref{fig1} (b), 
where the transfer energies are determined as 
$t_{1}=161.61$, $t_{2}=-157.29$, $t_{3}=-11.44$, $t_{4}=21.26$, 
and $t_{5}=-2.81$ [meV]. 
As seen in Figure \ref{fig1} (a), 
there are two band dispersions near the Fermi level, 
separated by a band gap.
The band width is estimated to be 1.28 eV, and 
the maximum value of the band gap is around 0.13 eV. 

\section{Boltzmann's equation approach}

Using the Boltzmann's equation 
approach, the thermopower is given by 
\begin{eqnarray}
 {\bf S}=\frac{1}{eT} {\bf K}_{0}^{-1} {\bf K}_{1}, 
 \label{Seebeck}
\end{eqnarray}
where $e(<0)$ is the electron charge, $T$ is the temperature, 
tensors ${\bf K}_0$ and ${\bf K}_1$ are given by
\begin{eqnarray}
 {\bf K}_{n}=\sum_{\textbf{\textit k}}
  \tau\left( \textbf{\textit k} \right)
  \textbf{\textit v}\left( \textbf{\textit k} \right)
  \textbf{\textit v}\left( \textbf{\textit k} \right)
  \left\{ 
   -\frac{\partial f\left( \varepsilon_{\textbf{\textit k}} \right)}
   {\partial \varepsilon_{\textbf{\textit k}} }
  \right\}
  \left\{ \varepsilon_{\textbf{\textit k}} - \mu \right\}^{n}, 
 \label{Kn}
\end{eqnarray}
Here, $\varepsilon \left( \textbf{\textit k} \right)$ 
is the band dispersion, 
$\textbf{\textit v}\left( \textbf{\textit k} \right) = 
\nabla_{\textbf{\textit k}} 
\varepsilon\left( \textbf{\textit k} \right)$ 
is the group velocity, 
$\tau\left( \textbf{\textit k} \right)$ is the quasiparticle lifetime,  
$f\left( \varepsilon \right)$ is the Fermi distribution function,  
and $\mu$ is the chemical potential. 
Hereafter, we simply refer to $({\textbf K}_n)_{xx}$ as $K_n$, and 
$S_{xx}=(1/eT)\dot(K_1/K_0)$ (for diagonal ${\textbf K}_0$) as $S$. 
Using $K_0$, conductivity can be given as 
$\sigma_{xx}=e^2K_0\equiv\sigma$.
Roughly speaking for a constant $\tau$, 
\begin{eqnarray}
K_0\sim\Sigma'(v_A^2+v_B^2), \,\,\,\,
K_1\sim(k_BT)\Sigma'(v_B^2-v_A^2)
\end{eqnarray}
(apart from a constant factor) stand, 
where $\Sigma'$ is a summation over the states in the range of 
$|\varepsilon(\textbf{\textit k})-\mu|<\sim k_BT$, 
and $v_A$ and $v_B$ are typical velocities for the states 
above and below $\mu$, respectively.
In usual metals, where $v_A\sim v_B$, 
the positive and negative 
contributions in $K_1$ nearly cancel out to  
result in a small $S$.
Now, let us consider a band 
that has a somewhat flat portion at the top (or the bottom), 
which sharply bends into a highly dispersive portion below (above).
We will refer to this band structure as the 
``pudding mold'' type \cite{Kuroki-Arita-JPSJ-76-083707}.
For this type of band with $\mu$ sitting near the 
bending point, $v^2_A\gg v^2_B$ holds for high enough temperature, 
so that the cancellation in $K_1$ is less effective, 
resulting in $|K_1|\sim(k_BT)\Sigma'v_A^2$ and $K_0\sim\Sigma'v_A^2$, 
and thus large $|S|\sim O(k_B/|e|)\sim O(100)\mu$V/K. 
An important point for this type of band is that 
the large $v_A$ and  the large 
Fermi surface result not only in large  
$|\frac{K_1}{K_0}|\propto S$ but also 
in large $K_0\propto\sigma$ as well, 
being able to give a large power factor $S^2\sigma$.
The pudding mold type band scenario well explains the coexistence of 
large thermopower and metallic conductivity observed in Na$_{x}$CoO$_{2}$ 
\cite{Terasaki-Sasago-PRB-56-R12685}. 

Now, if we look at the band structure of the $\tau$-type conductor 
from this viewpoint, the lower band has a nearly flat portion at the top, 
while the upper band has a nearly flat portion at the bottom.
Therefore, the band structure can be considered 
as ``inverted pudding mold type band'' on top of 
a ``pudding mold type band''.

\section{Thermopower} 
\label{Thermopower} 
%
Using the Boltzmann's equation approach and 
assuming a rigid band, 
we have calculated the temperature dependence of the Seebeck coefficient 
for the band filling $n=1.0625$ and $n=1.125$, which corresponds to 
the anion content of $y=0.875$ and $y=0.75$, respectively.
 \begin{figure}[h]
\begin{center}
  \includegraphics[width=7.0cm]{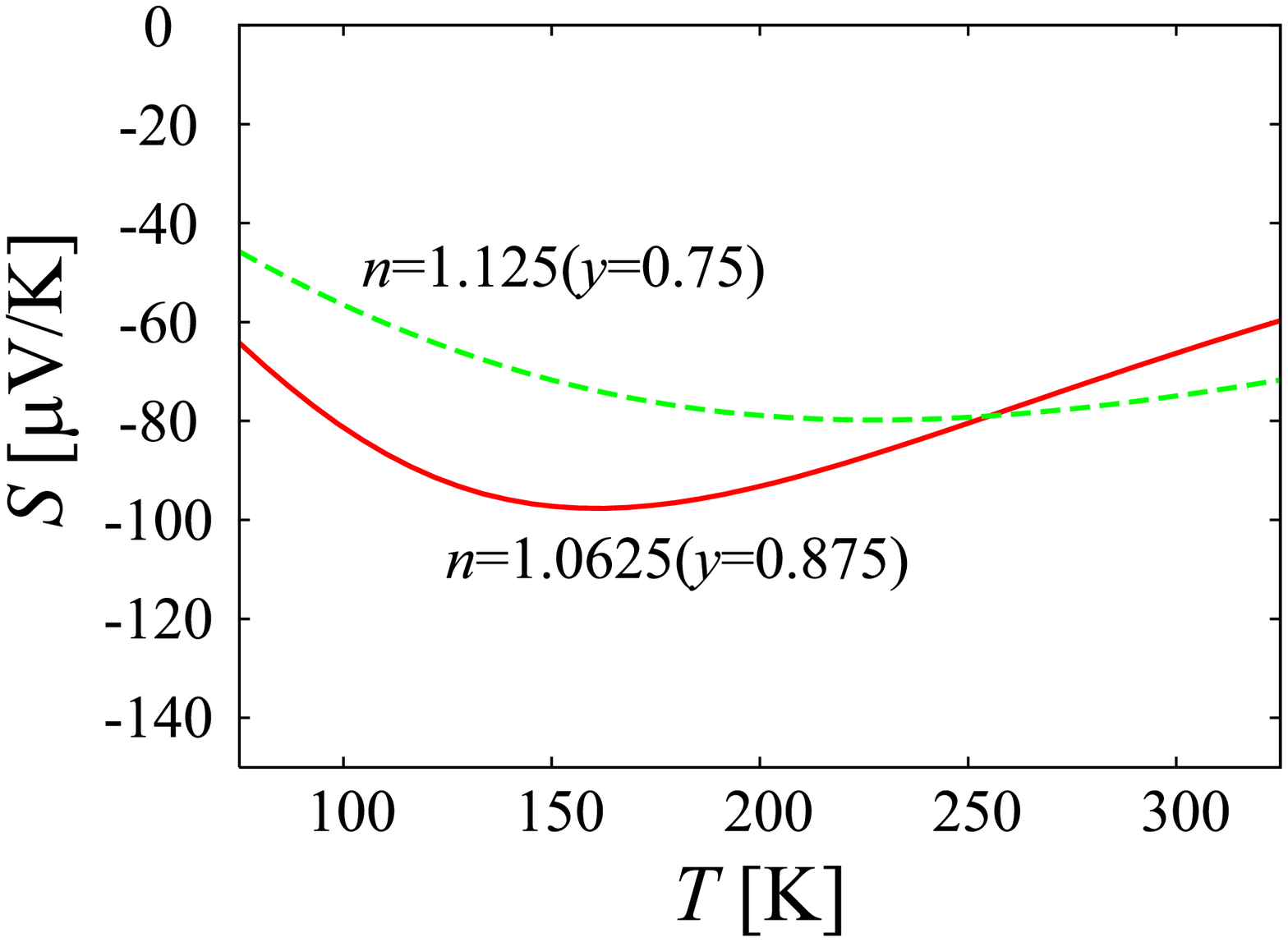}
\end{center}
{Fig.2  The calculated Seebeck coefficient as functions of the temperature 
for the band filling of $n=1.0625$ (solid red) and $1.125$ (dashed green),
corresponding to $y=0.875$ and $0.75$, respectively.
  \label{fig2}}
 \end{figure}
%
%
Although the maximum absolute value of the calculated $S$ 
is somewhat smaller than the experimental one 
\cite{Yoshino-Papavassiliou-JTAC}, 
the overall tendency of the temperature dependence is well reproduced 
by the calculation when we adopt $y=0.875$. 

The reason for the appearance of the maximum value of $S$ 
can be understood as follows. 
In the low temperature regime, due to the presence of the 
band gap, only the upper band is effective, which gives a negative Seebeck 
coefficient. The Seebeck coefficient in this temperature regime 
can be large (despite the metallic conductivitiy)  
because of the inverted 
pudding mold shape of the upper band.
At higher temperatures, the lower band, which 
gives a positive contribution to $S$, becomes effective, 
and this suppresses the absolute value, resulting in a
maximum value at a certain temperature.

\section{Conclusion}
\label{Conclusion}

We have studied the origin of the large thermopower in the $\tau$-type 
organic conductor. 
We have performed an {\it ab initio} band calculation by using WIEN2K, 
and made a tight binding model fit to the \textit{ab initio} band. 
Using this tight binding model, 
we have calculated the Seebeck coefficient which nicely reproduces 
the experimental results.
We conclude that the peculiar band structure consisting of 
pudding + inverted pudding pudding mold type bands is the origin 
of the large thermopower as well as the appearance of the 
maximum value at a certain temperature. A more detailed analysis 
(such as on the molecule dependence) 
is now underway\cite{Aizawatau}.
\ \\

\noindent
{\bf Acknowledgement} 

KK acknowledges H. Mori for pointing out the possible relevance of the 
peculiar band structure of $\tau$-type conductors to the large Seebeck 
coefficient. Numerical calculations were performed
at the facilities of the Supercomputer Center,
ISSP, University of Tokyo.
This study has been supported by 
Grants-in-Aid for Scientific Research from the Ministry of Education, 
Culture, Sports, Science and Technology of Japan, and from the Japan 
Society for the Promotion of Science.


\clearpage

\end{document}